# New research paradigms and agenda of human factors science in the intelligence era

XU Wei[1], GAO Zaifeng[2], GE Liezhong[1]

([1] Center for Psychological Sciences; [2] Department of Psychology, Zhejiang University, Hangzhou 310058, China)

## Abstract

This paper first proposes the innovative concept of "human factors science" to characterize engineering psychology, human factors engineering, ergonomics, human-computer interaction, and other similar fields. Although the perspectives in these fields differ, they share a common goal: optimizing the human-machine relationship by applying a "human-centered design" approach. AI technology has brought in new characteristics, and our recent research reveals that the human-machine relationship presents a trans-era evolution from "human-machine interaction" to "human-AI teaming." These changes have raised questions and challenges for human factors science, compelling us to re-examine current research paradigms and agendas.

In this context, this paper reviews and discusses the implications of the following three conceptual frameworks that we recently proposed to enrich the research paradigms for human factors science. (1) human-AI joint cognitive systems: This model differs from the traditional human-computer interaction paradigm and regards an intelligent system as a cognitive agent with a certain level of cognitive capabilities. Thus, a human-AI system can be characterized as a joint cognitive system in which two cognitive agents (human and intelligent agents) work as teammates for collaboration. (2) human-AI joint cognitive ecosystems: An intelligent ecosystem with multiple human-AI systems can be represented as a human-AI joint cognitive ecosystem. The overall system performance of the intelligent ecosystem depends on optimal cooperation and design across the multiple human-AI systems. (3) intelligent sociotechnical systems (iSTS): human-AI systems are designed, developed, and deployed in an iSTS environment. From a macro perspective, iSTS focuses on the interdependency between the technical and social subsystems. The successful design, development, and deployment of a human-AI system within an iSTS environment depends on the synergistic optimization between the two subsystems.

This paper further enhances these frameworks from the research paradigm perspective. We propose three new research paradigms for human factors science in the intelligence ear: human-AI joint cognitive systems, human-AI joint cognitive ecosystems, and intelligent sociotechnical systems, enabling comprehensive human factors science solutions for AI-based intelligent systems. Further analyses show that the three new research paradigms will benefit future research in human factors science. Furthermore, this paper looks forward to the future research agenda of human factors science from three aspects: "human-AI interaction", "intelligent human-machine interface", and "human-AI teaming". We believe the proposed research paradigms and the future research agenda will mutually promote each other, further advancing human factors science in the intelligence era.

Corresponding authors: Xu Wei, email: xuwei11@zju.edu.cn; Gao Zaifeng, email: zaifengg@zju.edu.cn

The original article is in Chinese. The Chinese version shall always prevail in case of any discrepancy or inconsistency between the Chinese version and its English translation. The English version of the article was translated and polished by Pan Jiawei, email: 3200104850@zju.edu.cn

# 1 Introduction

Engineering psychology, human factors engineering, and ergonomics emerged during World War II. Although each has unique perspectives and focuses, they share the "human-centered" design philosophy. For example, engineering psychology provides psychological principles, methods, and empirical evidence for optimizing the design of human-machine systems from the perspective of human cognitive information processing (Wickens et al., 2021; Sun et al., 2011; Xu & Zhu, 1989). Human factors engineering and ergonomics contribute human factors and ergonomics design principles, methods, and data for optimizing human-machine-environment relationships from a design perspective (Sanders & McCormick, 1993; Xu & Chen, 2012, 2013, 2014; Xu & Zhu, 1990; Xu, 2007). As we entered the computer era, interactions between humans and computers (including products based on computing technology) brought forth new human factors issues, driving the emergence and development of fields such as human-computer interaction (HCI) and user experience (Norman & Kirakowski, 2017; Norman & Draper, 1986; Xu et al., 1999; Xu, 2003, 2005, 2007). Driven by the shared "human-centered" design philosophy, these related fields aim to optimize interactions between humans, machines, and environments, ensuring safety, efficiency, and user satisfaction for systems. Therefore, these fields are collectively referred to as Human Factors Science.

This paper introduces the innovative concept of "human factors science" for the first time. This concept highlights the shared research ideologies, objects, and purposes that engineering psychology, human factors engineering, ergonomics, human-computer interaction, and user experience adopt around the core element of "human" in their respective research and applications. Similar to the relationship between cognitive science and cognitive psychology, the concept of human factors science is a higher-level field relative to these individual fields. Many research and application problems encountered in practice sometimes cannot be simply attributed to a specific field within these related areas. It often requires exploring comprehensive solutions from an interdisciplinary perspective. Therefore, the concept of the "human factors science" discipline facilitates the exploration of comprehensive solutions at an elevated system level, fostering collaboration between these related fields to achieve common goals.

In the intelligence era, intelligent autonomous systems based on AI technologies bring forth a series of new features and problems (Kaber, 2018; Xu, 2020). This will undoubtedly pose new requirements for the research ideologies and paradigms of human factors science, requiring a systematic review to ensure human factors science can make more effective contributions to the optimized design of new intelligent technologies. This paper defines the research paradigm of a field as a lens that frames the perspective of research and determines the research scope, focus, and corresponding methods. Starting from the overall perspective of human factors science, this paper provides a comprehensive and systematic assessment of the research paradigms and focuses on related fields within human factors science.

In the past five years, we have proposed a series of concepts, models, and frameworks to address issues related to the research ideologies and objects of human factors science in the era of intelligence. These include the "human-centered AI" concept, a novel human-machine relationship based on "human-AI teaming," and the new interdisciplinary research field of "human-AI interaction". Based on this work, we have further undertaken new explorations and proposed a series of conceptual models and frameworks in terms of research paradigms and agenda in the intelligence era (Xu et al., 2019; Xu, 2019, 2021; Xu & Ge, 2018, 2020; Xu et al., 2021; Xu, 2019a, 2022, 2022a, 2022b, 2022c; Xu et al., 2022; Xu & Dainoff, 2023; Ozmen Garibay et al., 2023; Xu & Gao, 2023; Gao et al., 2023). Our previous work brings new considerations to the research paradigms and focuses on human factors science. However, the research of human factors science has not yet systematically conducted work in this regard. By reviewing the research paradigms and focuses of human factors science, this paper further enhances the new concepts and frameworks we have proposed, addressing the following scientific questions: What research paradigms should be adopted for effectively supporting future research in human factors science in the intelligence era? What are the future research focuses for human factors science in the intelligence era? This paper aims to provide new perspectives for the next steps in human factors science research.

# 2 Human Factors Science: Evolution of Research Objects and Paradigms Across Eras

## 2.1 Evolution of the Research Objects of Human Factors Science

The objects of human factors science is the human-machine relationship. In the era of computers, machines played the role of auxiliary tools in the interaction between humans and non-intelligent computing systems. In the intelligence era, the interaction between humans and intelligent systems essentially involves interactions with intelligent agents within the intelligent system. Leveraging intelligent technology, these intelligent agents can exhibit unique autonomous features, possessing cognitive capabilities similar to humans (e.g., perception, learning, reasoning, etc.). In unforeseen design scenarios, they can autonomously perform tasks that were previously beyond the scope of traditional automation technology (Kaber, 2018; Madni & Madni, 2018; Xu, 2020). Therefore, intelligent systems can evolve from auxiliary tools supporting human operations to team members cooperating with human teammates, playing a dual role of "auxiliary tool + human-machine collaborative teammate". This forms a new type of human-machine relationship termed "Human-AI Teaming" (Brill et al., 2018; Xu & Ge, 2020). This novel collaborative relationship, "human-AI teaming", imbues the human-machine relationship with new meanings, leading to a cross-generational evolution of human-machine relationships (Xu & Ge, 2020). It also introduces a new perspective for human factors science research, necessitating a reexamination of the previous research paradigms focusing on non-intelligent technology.

## 2.2 Exploration of Research Paradigms in Human Factors Science

Throughout history, the research paradigms of human factors science have continually expanded with the emergence of new technologies. This expansion has elevated the methodological aspects of human factors, broadened the scope of disciplinary research, and increased the depth of problem-solving, thereby propelling the continuous development of human factors science.

In its early stages of development, traditional human factors science (such as ergonomics) focused on physical tasks, human-machine interfaces, and other physical characteristics. Research primarily utilized methods such as task analysis and time-task analysis to achieve a rational allocation of human-machine functions and tasks, aiming to maximize the efficiency of human-machine systems (Sanders & McCormick, 1993; Gardner et al., 1995).

Since the advent of the computer era, the research paradigm of human factors science has been essentially constructed based on cognitive theories of information processing. However, this paradigm has exhibited different orientations at different stages of the development of human factors science. In the computer era, human factors science (such as engineering psychology and cognitive ergonomics) started from information processing mechanisms and delved into the psychological aspects of human activities. It investigated the relationships between human performance and psychological activities (e.g., perception, attention, memory, and decision-making) under a human-machine operating environment, optimizing the design of human-machine systems (Wickens et al., 2021; Sun et al., 2011). To address human factors issues in human-computer interactions, human factors science (such as human-computer interaction and user experience) adopted methods based on the "user-centered" design philosophy. The work in these areas involves building user mental models, situational awareness models, cognitive models of human-computer interaction, conceptual models of human-machine interfaces, and conducting usability tests based on psychological methods to validate and develop interactive products that align with user needs and experiences (Nielsen, 1994; Finstad et al., 2009; Xu, 2023).

However, the research of human factors science typically emphasizes the external behavioral aspects of humans, investigating cognitive activities and performance through objective work performance and subjective evaluation methods. To overcome the limitations of this approach, human factors science further incorporated measurement techniques from cognitive neuroscience (such as EEG, fMRI, etc.), giving rise to neuroergonomics. This field delves into the internal neural activity of human cognitive processing, exploring the neural mechanisms of cognitive processing in human-machine interactions (Parasuraman & Rizzo, 2006), thus providing more objective empirical means for human factors science (Dehais et al., 2020).

In the intelligence era, the new features of intelligent technology and the cross-generational evolution of the research ideologies and objects (human-machine relationship) of human factors science inevitably lead to new considerations for the research paradigms of human factors science. However, the human factors science community has not yet systematically undertaken the relative work. Since 2021, we have conducted new explorations into the research paradigms of human factors science in the intelligence era. In this paper, we summarized the conceptual models and frameworks we proposed, including human-AI joint cognitive systems, human-AI joint cognitive ecosystems, and intelligent sociotechnical systems (Xu, 2022a, 2022b, 2022c; Xu et al., 2022; Xu & Dainoff, 2023; Ozmen Garibay et al., 2023; Xu & Gao, 2023; Gao et al., 2023). Furthermore, this paper enhances these conceptual frameworks from the perspective of research paradigms of human factors science. Then, we analyze the application significance of these research paradigms through an application example of human-vehicle co-driving, and finally provide prospects for future research.

# 3 New Research Paradigms of Human Factors Science in the Intelligence Era

## 3.1 Human-AI Joint Cognitive Systems

### 3.1.1 The Conceptual Framework of Human-AI Joint Cognitive Systems

Human-machine interaction models are a crucial component of research in human factors science. Researchers commonly adopt models of human information processing, such as MHP, GOMS, SOAR, ACT-R, EPIC, to construct human-machine interaction models (Wang et al., 2020). These models primarily focus on non-intelligent human-machine interactions, where machines are considered merely tools, without accounting for the potential collaborative relationship between humans and intelligent systems, thus rendering them ineffective for handling tasks involving human-AI interaction (Xu & Ge, 2020). Currently, there is a consensus regarding the collaborative relationships between humans and intelligent systems (e.g., NASEM, 2021; Caldwell et al., 2022). However, there is a lack of interaction models for human-AI teaming. In light of this, considering the emergence of autonomy based on intelligent technologies and the "human-AI teaming" metaphor, and drawing upon theories such as Joint Cognitive Systems (Hollnagel & Woods, 2005), Situation Awareness (Endsley, 1995), and Intelligent Agent theory (Wooldridge & Jennings, 1995; Wang & Zhang, 2020), we propose, for the first time, a conceptual model for human-AI teaming in intelligent human-machine systems, termed the Human-AI Joint Cognitive Systems (Xu, 2022b; Xu & Gao, 2023) (see Figure 1).

As depicted in Figure 1, unlike traditional human-machine interaction systems, this model views intelligent systems (including one or more intelligent agents) as cognitive agents capable of performing specific cognitive information-processing tasks. Consequently, a human-AI system can be conceptualized as a joint cognitive system where two cognitive agents (humans and intelligent agents) collaborate. As a teammate collaborating with human users, intelligent systems engage in bidirectional proactive interaction and collaboration with human users using natural and effective human-machine interaction modalities such as speech, gestures, and expressions. In specific scenarios, intelligent systems autonomously perceive, recognize, learn, and infer user states (cognitive, physiological, intent, emotions, etc.) and environmental context, subsequently executing corresponding autonomous actions (Kaber, 2018; Xu, 2019b; Xu, 2021).

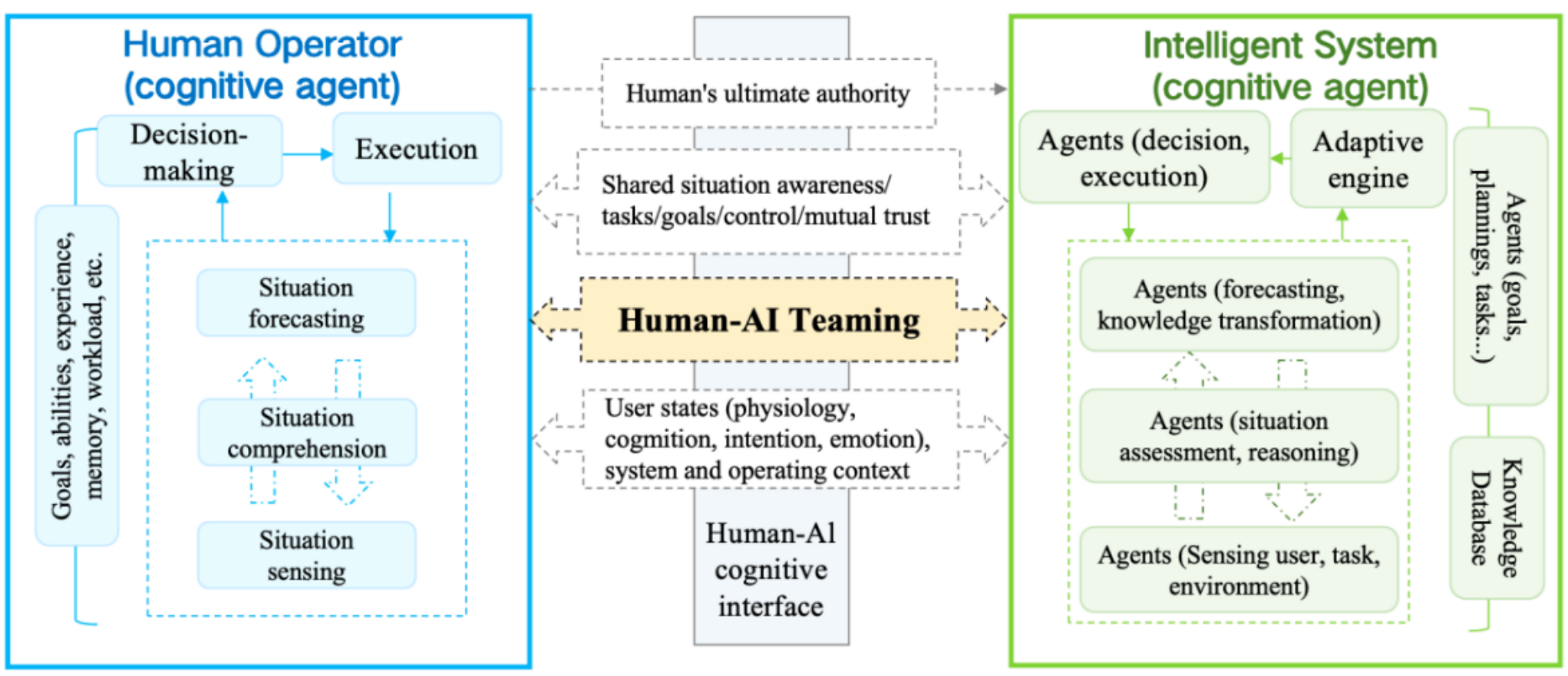

Figure 1 A conceptual framework of the human-AI joint cognitive systems

This framework employs Endsley's Situation Awareness theory to characterize the information processing mechanisms of human users and intelligent agents (Endsley, 1995, 2015). It encompasses interactions between situation awareness and cognitive factors such as memory, experience, and knowledge. It also incorporates information processing mechanisms based on data-driven (understanding and predicting situations based on sensory data) and goal-driven (validating sensory data based on goals, current understanding, and predictions) approaches. With a dynamic feedback and feedforward loop mechanism for information collection and subsequent responses, human users can perceive environmental situations to update acquired information. As illustrated in Figure 1 (the right side), this model represents the information processing mechanism for the cognitive machine agents of intelligent systems, which is similar to human agents.

The conceptual framework represented in Figure 1 provides a new research paradigm and perspective for human factors science, as characterized by several novel features as follows:

(1). A paradigm shift in research based on machine cognitive agents: Unlike traditional human factors science paradigms that consider machines as tools assisting human tasks, this model represents machine intelligent agents as cognitive agents collaborating with humans. This facilitates human factors science to enhance human-machine system performance by studying the cognitive behavior of intelligent agents and their collaborative behaviors with humans.

(2). A new collaborative research paradigm based on "human-AI teaming": Different from the traditional human factors science paradigm of "human-machine interaction", this design paradigm characterizes the human-machine relationship as a joint cognitive system between the two cognitive agents, exploring ways to enhance human-machine system performance through optimized human-machine collaboration.

(3). The design philosophy of "human-centered AI": Human users are the leaders of this human-AI collaborative team and serve as the ultimate decision-makers in emergencies.

(4). Bidirectional proactive state recognition in human-AI interaction: Different from traditional "stimulus-response" unidirectional human-machine interaction, this model emphasizes bidirectional proactive state recognition. Intelligent systems can actively monitor and identify users' physiological, cognitive, behavioral, intent, and emotional states through sensing systems, while human users obtain optimal situation awareness through multimodal human-machine interfaces.

(5). Human-machine intelligence complementarity: As a joint cognitive system, system performance depends not only on the performance of individual system components but also on human-machine hybrid intelligence, maximizing human-machine collaboration and overall system performance through the complementarity of human and machine intelligence.

(6). Adaptive intelligent human-machine interaction: Emphasizing adaptive mechanisms of intelligent systems, the model suggests that intelligent agents can produce appropriate adaptive system outputs in scenarios where the design cannot predict based on sensed recognition and inference of user states and environmental context. Human users, in turn, adaptively adjust their interaction behaviors based on situation awareness, tasks, goals, etc.

(7). Collabotation-based cognitive interfaces for human-machine interaction: The model emphasizes the construction of human-machine collaboration-based cognitive interfaces based on multimodal interaction technologies to support human-machine collaboration. This includes support for bidirectional situation awareness, human-machine mutual trust, shared decision-making and control, social interaction, emotional interaction, etc.

#### 3.1.2 Application Analysis

Autonomous vehicles based on intelligent technology represent a typical human-AI teaming system. Despite the involvement of human factors science professionals in developing autonomous vehicles, frequent accidents have prompted the exploration of new design approaches (NTSB, 2017; Endsley, 2018; Xu, 2020). SAE (2019) classifies the automation of autonomous vehicles into five levels (L1~L5). For a considerable time, human-vehicle co-driving will be the norm (Zong et al., 2021). In high levels of automated driving, adopting a research paradigm based on the human-AI joint cognitive systems framework, human drivers and in-vehicle intelligent systems are two cognitive agents capable of performing certain cognitive information processing tasks. Co-driving in autonomous vehicles represents a human-AI joint cognitive system (Xu, 2020; Xu, 2021; Gao et al., 2023). In-vehicle intelligent systems (agents) equipped with intelligent sensing technologies can perform specific cognitive tasks like sensing, recognizing, learning, and inferring based on human driver states and environmental context, facilitating effective human-vehicle co-driving. Therefore, solutions for human-vehicle co-driving can be explored in several dimensions:

(1). New design paradigm based on human-AI collaboration: The current classification of automated driving levels emphasizes the role of in-vehicle intelligent systems as tools to assist or replace human drivers, which is a "technology-centered" design philosophy (SAE, 2019). The human-AI joint cognitive system framework advocates a human-centered and human-AI collaborative design approach, facilitating the establishment of collaborative relationships and dynamic allocation of human-machine functions in various driving scenarios. This approach will help design optimization through human-machine collaboration, mutual trust, shared situation awareness, shared control, and collaborative driving (Biondi et al., 2019; Xu, 2020; Gao et al., 2021). For example, future research needs to study the transfer of vehicle control between humans and machines in emergencies, ensuring humans have ultimate control (including remote control) (Fridman, 2018); future work needs to explore under what conditions (such as the level of human-machine trust, driver states, and driving intentions) effective transfer of human-machine control can be achieved.

(2). Bidirectional recognition of human-vehicle states: The paradigm emphasizes bidirectional proactive state recognition between human and machine agents based on the monitoring and understanding of the human driver's state, driving behavior, and intentions by the machine agent, achieving effective human-machine collaboration For example, we need to study the modeling of driver intention to improve the accuracy of driver state monitoring; we need to study explainable AI to enhance human operator situation awareness.

(3). Hybrid intelligence design based on the "human-in-the-loop" mechanism: The "human-out-of-the-loop" issue is one of the significant contributors to accidents (NTSB, 2017; Endsley, 2018). Human-vehicle co-driving should be a "human-in-the-loop" hybrid intelligence system. Research should focus on effective "human-in-the-loop" designs at both the system level (such as "human-in-the-loop" control) and the biological level (such as brain-machine control). Future research also needs to investigate how to use human-machine complementarity to optimize system design, how to implement the "meaningful human control" design, and how to use in-vehicle "failure tracking systems" to track accountability for human-machine faults and identify opportunities for design improvement (Santoni et al., 2018).

(4). Autonomy-based design: As a "mobile" intelligent autonomous system, autonomous vehicles pose new demands on system design to handle emerging design features different from traditional automation technologies, such as potential indeterministic system outputs, explainable human-machine interfaces, and shared human-machine control (Xu, 2020). System design must adopt effective methods to meet these new requirements for driving safety.

(5). Collaboration-based cognitive interface design: The paradigm suggests future work to explore "collaboration-based cognitive interfaces", supporting collaborative human-vehicle co-driving. Future work needs to investigate effective design metaphors, paradigms, and cognitive architectures for human-machine interfaces, such as in-vehicle ecological user interfaces (Burns & Hajdukiewicz, 2004).

### 3.1.3 Research Prospects

To enrich the research paradigm of Human-AI Joint Cognitive Systems, future research should consider the following aspects: (1). Conduct research on cognitive abilities, behavioral evolution, and other topics for machine agents (Zhu et al., 2020), investigating how human factors science can contribute to optimizing the design of machine intelligent agents and enhancing the collaboration of human-AI teaming. (2). Conduct theoretical research on human-AI teaming, developing theories, methods, and performance evaluation systems for human-AI teaming by leveraging existing human-human teaming theories (Madhavan & Wiegmann, 2007; Kaber, 2018). (3). Construct models for human-machine collaboration-based cognitive systems, including shared and team situation awareness, mutual trust, and shared decision-making and control (Xu, 2020; Gao et al., 2023). (4). Develop effective cognitive models for recognizing user states (e.g., behaviors, intentions, and emotions), conducting research on effective knowledge representation and knowledge graphs to support computational modeling. (5). Explore adaptive optimization design based on assessments of user, system, and environmental state, and utilize intelligent agents' proactive predictive capabilities to assist human teammates in achieving proactive, adaptive human-machine interaction. (6). Build effective paradigms and models of collaboration-based cognitive interface design to support effective human-machine collaboration.

### 3.2 Human-AI Joint Cognitive Ecosystems

#### 3.2.1 The Conceptual Framework of Human-AI Joint Cognitive Ecosystems

Research in human factors science typically focuses on individual human-machine systems, and research on human-AI joint cognitive systems mainly addresses the collaboration between humans and individual intelligent systems. However, intelligent technology is not merely a product or system but an ecosystem that encompasses technological transformation, system evolution, operational innovation, and organizational adaptation across intelligent systems (Liu et al., 2023). For instance, human-vehicle co-driving for individual autonomous vehicles represents just one facet of a human-AI joint cognitive system. The entire intelligent co-driving ecosystem includes interactions and collaboration between humans and vehicles, vehicles and other vehicles, and vehicles and the intelligent traffic environment by leveraging technologies like intelligent vehicular networks and intelligent traffic systems. These multi-agent systems constitute a human-AI joint cognitive ecosystem. The interactions and collaboration among these multiple human-AI joint cognitive systems directly influence the driving safety of individual vehicles. Therefore, the research paradigms in human factors science need to move beyond focusing on a single human-AI system and adopt a more holistic approach to consider systematic solutions based on an ecosystemic research paradigm.

Currently, research on multi-agent systems primarily emphasizes engineering and technological aspects (Dorri et al., 2018; Allenby, 2021; Ali et al., 2021), such as distributed multi-agent systems and human-machine-object fusion swarm intelligence computing (Xie & Xie, 2021; Guo & Yu, 2021). However, there is a lack of studies considering system design from the perspective of human factors science. We propose an ecosystemic research paradigm, drawing inspiration from theories such as joint cognitive systems (Hollnagel & Woods, 2005), multi-agent systems theory (Dorri et al., 2018; Le Page & Bousquet, 2004), and multi-agent ecosystem thinking (IDC, 2020; Allenby, 2021; Ali et al., 2021). For the first time, we characterize an intelligent ecosystem as a human-AI joint cognitive ecosystem (see Figure 2) (Xu, 2022a). As illustrated in Figure 2, a multi-agent ecosystem can be represented as a joint cognitive ecosystem comprising a series of human-AI joint cognitive systems (such as smart cities and intelligent transportation).

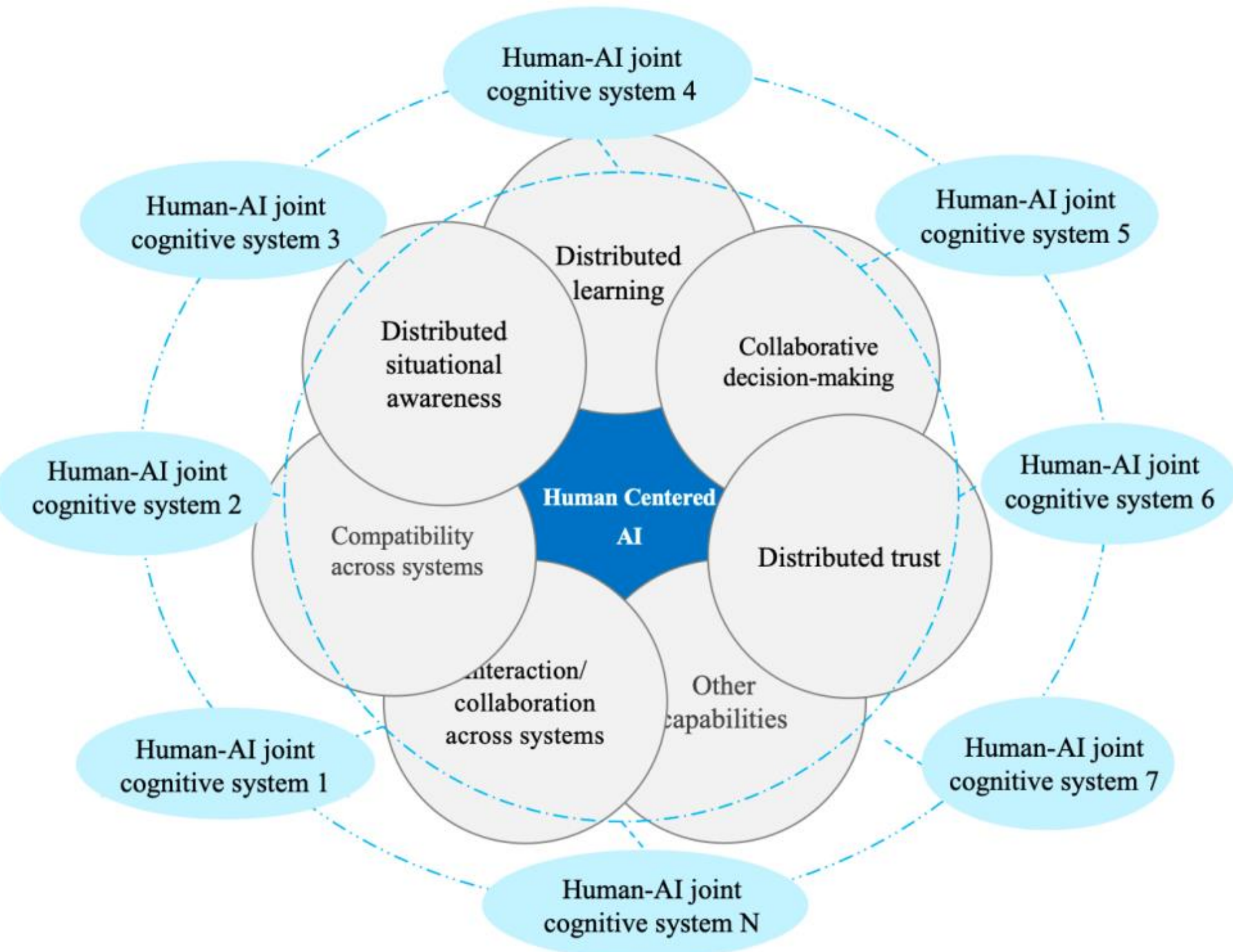

Figure 2 A conceptual framework of human-AI joint cognitive ecosystem

The conceptual framework depicted in Figure 2 provides a novel research paradigm for human factors science from an ecosystemic perspective, reflecting several new features and research directions as follows:

(1). A novel ecosystem-based paradigm: A multi-agent ecosystem is represented as a human-AI joint cognitive ecosystem comprising a series of interacting human-AI joint cognitive systems. The overall optimization design of intelligent ecosystems needs to consider the interactions and collaboration among these systems. This paradigm shift extends the focus of human factors science research from a "point" solution (individual human-machine system) to a "2-D plane" solution (cross-human-machine systems).

(2). "Human-centered AI" design philosophy: Humans hold a central position within intelligent ecosystems, emphasizing that the design, development, and deployment of intelligent ecosystems must prioritize human needs, values, intelligence, capabilities, and roles, for example, setting human-AI decision-making authority across systems, establishing trust ecosystems, and resolving conflicts between intelligent systems that may be built based on different cultures, societies, and ethics. The goal is to ensure humans have ultimate control over the entire human-AI ecosystem for overall safety.

(3). Distributed collaboration: Based on the ecosystemic paradigm, optimal performance of the overall ecosystem can be achieved through effective distributed human-AI interactions and collaboration across intelligent systems. The design of distributed systems needs to consider distributed and shared cognitive-enhanced learning, situation awareness, human-AI emotional interaction, human-AI trust, human-AI information processing, human-AI cognitive learning, collaborative decision-making, and social interaction across intelligent systems, enhancing the overall collaboration of the human-AI ecosystem.

(4). System learning and evolution: System design should emphasize learning and evolution within human-AI ecosystems. Leveraging features of learning and evolution in both human and AI systems, the emphasis is on continuous evolution and optimization across human-AI systems. This involves distributed cognitive-enhanced learning, cross-agent and cross-task collective intelligence knowledge transfer, self-organization, adaptive collaboration, etc., adapting elements within the system to dynamic and complex application scenarios.

### 3.2.2 Application Analysis

The conceptual framework of the human-AI joint cognitive ecosystem has been preliminarily applied to analyze scenarios such as human-vehicle co-driving in autonomous vehicles and single pilot operations (SPO) in large commercial aircraft cockpits (Xu, 2022a; Xu et al., 2021). Taking human-vehicle co-driving as an example, the safety of human-vehicle co-driving relies not only on the system design at the individual vehicle level but also on achieving collaboration among various cognitive agents in humans, vehicles, roads, and the intelligent traffic environment. This is facilitated through effective information exchange between vehicle, road, and cloud, optimizing the system design for safe driving and decision-making of the entire human-vehicle co-driving ecosystem (Tan et al., 2020). Figure 3 illustrates the human-AI joint cognitive ecosystem for human-vehicle co-driving, providing new perspectives for human factors science research in developing human-vehicle co-driving solutions.

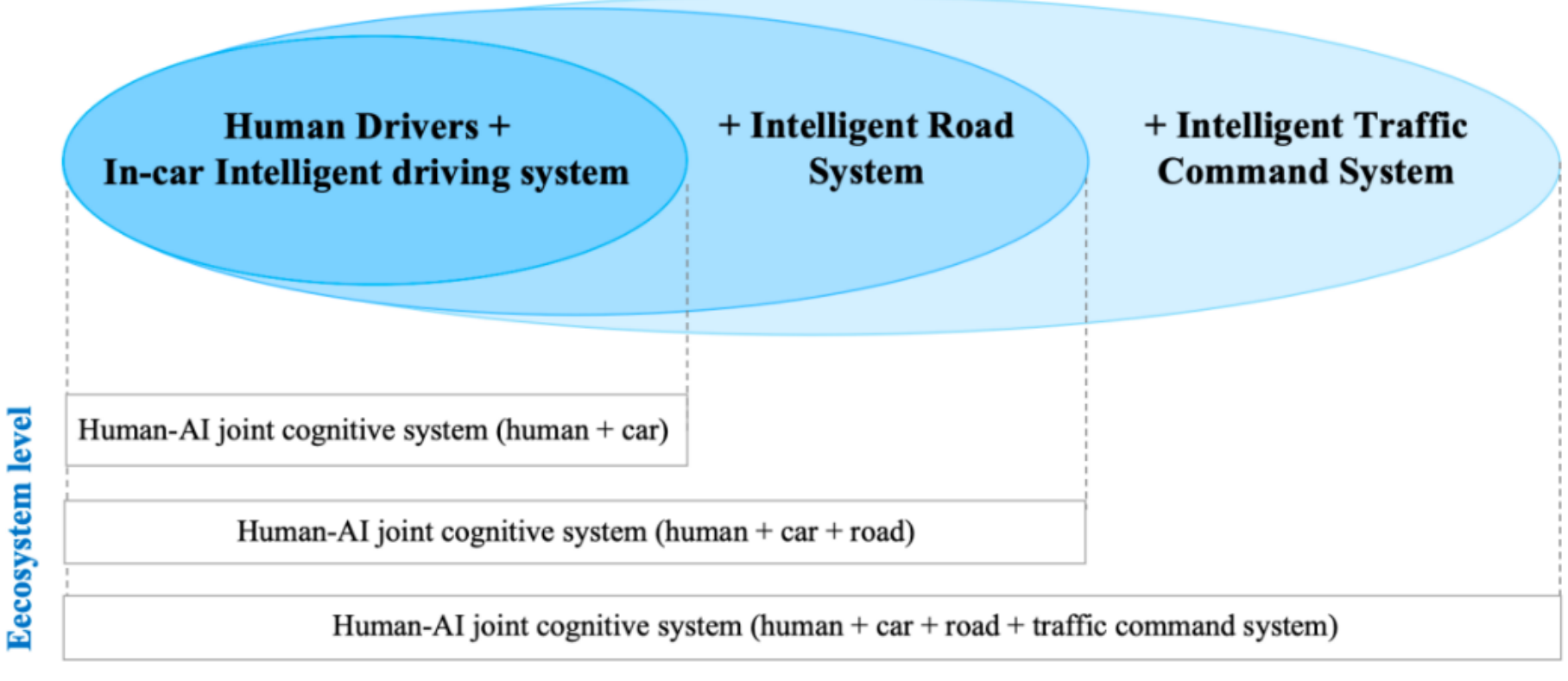

Figure 3 A schematic diagram of the human-vehicle co-driving cognitive ecosystem

Firstly, we should adopt systematic design approaches for human-vehicle co-driving. The design philosophy must transcend the limitations of individual vehicle-centered perspectives, such as "human driver + intelligent autonomous vehicle". Such a narrow focus cannot guarantee the optimized design and safe operations of the entire joint cognitive ecosystem. As an extensive engineering endeavor of the human-AI joint cognitive ecosystem, the safety of the co-driving ecosystem (human-vehicle-road-traffic-societal system) depends on the effective collaboration and optimized design among all human-AI joint cognitive systems within the ecosystem.

Secondly, system design needs to achieve the "human-centered AI" design philosophy at the human-AI joint cognitive ecosystem level, ensuring that humans retain ultimate control over autonomous vehicles. For example, in the event of a vehicle losing control (due to system failures, hacking attacks, or incapacitation of the driver), the in-car intelligent system should initiate intelligent emergency protocols to exit the current uncontrollable scenario, protecting humans (including drivers, passengers, pedestrians, and other vehicles), simultaneously, as one of the redundancy system safety measures within the human-AI joint cognitive ecosystem, operators at urban intelligent traffic control centers or intelligent vehicle operation centers should be capable of remotely taking over malfunctioning vehicles (e.g., "5G cloud chauffeur"). This involves coordinating and directing other autonomous vehicles on the road to ensure the overall safety of the intelligent road system. This ecosystem design must consider human capabilities, providing effective situation awareness (states of human, vehicle system, and driving environment) for all members in the system, ensuring real-time allocation of driving permissions and responsibilities, and guaranteeing that humans have ultimate control over the human-vehicle co-driving ecosystem in emergencies.

Thirdly, system design needs to achieve system safety and optimization design for human-vehicle co-driving at the ecosystem level. From the ecosystem perspective, system design must ensure the realization of co-learning, co-evolution, and co-adaptive capabilities of all subsystems within the human-vehicle co-driving ecosystem. From a collaborative cognitive perspective, system design needs to ensure effective interaction among all subsystems in the human-vehicle co-driving ecosystem, including human-human, human-AI, cross-intelligent agents, and cross-subsystem interactions. This involves establishing effective compatibility and conflict resolution mechanisms (technical, traffic regulations, etc.) within the entire ecosystem to achieve effective collaboration.

Last but not least, we propose strategies for implementing human-vehicle co-driving systems from the perspective of human-AI joint cognitive ecosystems. Considering the conceptual framework illustrated in Figure 3, several potential implementation paths are outlined: (1). The "bottom-up" path: Starting from "human driver + intelligent autonomous vehicle" to "intelligent road systems" and "intelligent traffic control systems". (2). The "top-down" path: Beginning with "intelligent traffic systems" to "intelligent road systems" and "human driver + intelligent autonomous vehicle". (3). Hybrid or parallel paths: A combination of the "bottom-up" and "top-down" paths, representing potentially the safest and most effective implementation route.

### 3.2.3 Research Prospects

Enriching the conceptual framework of the human-AI joint cognitive ecosystems requires interdisciplinary collaboration. Future research should address the following fundamental issues: (1). Existing human factors science research focused on individual human-machine systems primarily employs empirical research methods. For research oriented towards multi-human-AI ecosystems, there is a need to explore methodologies that support this paradigm. This involves investigating the fundamental elements, characteristics, structures, development, and evolutionary processes of a joint cognitive ecosystem. (2). Exploration of fundamental theories is crucial, for example, the theory of human-machine behavioral coordination and symbiosis. This theory can investigate human-AI adaptability, self-organization, evolution, and other abilities based on changes in the human-machine environment, differences in human-machine perception capabilities and complementarity (Rahwan et al., 2019; Werfel et al., 2014; Guo & Yu, 2021). Also, there are theories on the evolution of intelligent agents in a group-distributed environment, encompassing features like collaborative fusion, competition, and knowledge transfer methods among intelligent agents within a group (Guo & Yu, 2021; Neftci & Averbeck, 2019). (3). Future work needs to explore the collaboration across multi-human-AI systems. This involves multi-agent human-machine interaction and collaboration mechanisms, performance evaluation systems, coordination, task allocation, team building, situation awareness, trust, task sharing, and decision control sharing among humans and multiple intelligent agents (Le Page & Bousquet, 2004; Dorri et al., 2018). Future work also needs to address the challenges related to compatibility, communication, mode transition, and the collaborative role of human users in interactions across different cultures and norms within multi-intelligent agent systems (ISO, 2020). (4). Efforts should be directed towards exploring human factors science processes and methodologies that can effectively influence the design of intelligent ecosystems. This involves investigating methods and procedures that can be applied to real-world scenarios to enhance the effectiveness and efficiency of intelligent ecosystem design.

### 3.3 Intelligent Sociotechnical Systems

#### 3.3.1 The Conceptual Framework of Intelligent Sociotechnical Systems

Sociotechnical Systems (STS) theory advocates for the coordination and optimization among subsystems, such as social, technical, and organizational components, aiming to achieve optimal system performance (Eason, 2008). Human factors science research usually focuses on the impact of the human-machine interface and the physical environment on the performance of human-machine systems, neglecting macro-level societal and organizational factors (Ge, Xu, & Song, 2022). Over the past two decades, STS theory has influenced the field of human factors science, driving the development of macroergonomics and related areas (Waterson et al., 2015). Given that intelligent systems operate within specific STS environments and intelligent technologies may have adverse effects on humans (such as privacy, ethics, decision-making authority, and skill growth), this negative impact has prompted human factors science research to consider the development and use of intelligent systems within the broader STS context (Stahl, 2021), leading to research specifically addressing intelligent systems (Asatiani et al., 2021). For instance, Steghofer et al. (2017) argue that the next generation of STS should be based on intelligent technologies, with the social subsystems of STS influenced by factors such as intelligent technologies and the potential decision uncertainty of AI. From the perspective of STS development processes and interdisciplinary collaboration, researchers have proposed approaches such as user-participatory design processes (Huang et al., 2019), the social-technical systems engineering (STSE) framework (Baxter & Sommerville, 2011), adaptive STS system architecture (Dalpiaz et al., 2013), and human factors science methodologies (Waterson et al., 2015; Hollnagel et al., 2006). The new field of "Human-AI Interaction", proposed by Xu et al. (2021), also considers the macroscopic STS environment.

Currently, there is no mature STS theory specifically tailored to intelligent technologies. Human factors science research should explore how to effectively develop and use intelligent systems in complex STS environments. To address this, Xu (2022c) analyzed the new features of the STS in the intelligence era. These features include aspects such as system composition, cognitive agents, human-machine relationships, user requirements, system decision-making and control, system learning capabilities, system design scope, organizational goals and requirements, system complexity, and openness. Based on these new features, we have proposed a conceptual framework for Intelligent Sociotechnical Systems (iSTS) (Xu, 2022c) (see Figure 4).

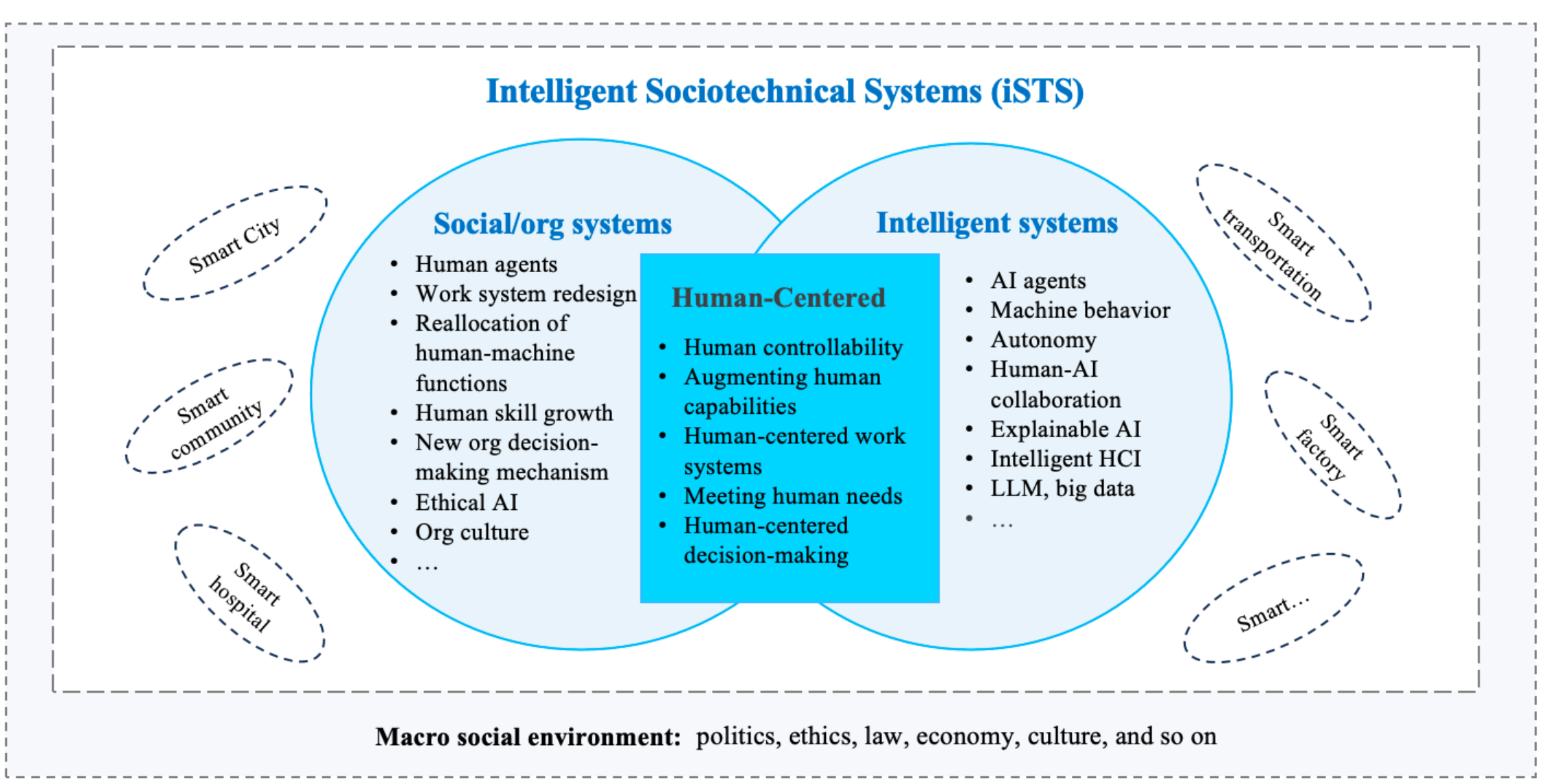

Figure 4 A conceptual framework of intelligent sociotechnical systems (iSTS)

As illustrated in Figure 4, iSTS inherits some fundamental characteristics from traditional STS theory. For example, iSTS has internally independent but interdependent technical and social subsystems, and the overall system performance relies on the collaborative optimization between these two subsystems (Badham et al., 2000). iSTS encompasses a macroscopic external environment and various forms of intelligent social structures. In contrast to the Human-AI Joint Cognitive Ecosystems, iSTS places greater emphasis on macroscopic and non-technical factors, including work system redesign, organization redesign, and intelligent decision-making.

The conceptual framework represented in Figure 4 provides a new paradigm for human factors science research from the sociotechnical perspective. This paradigm reflects several new features and research directions in human factors science:

(1). STS-based search paradigm: The development and utilization of any human-intelligence system (i.e., human-AI joint cognitive system) and intelligent ecosystem (i.e., human-AI joint cognitive ecosystem) occur within an intelligent STS environment. The optimization design and effective use of intelligent systems need to consider the interactions between technical and non-technical subsystems, showcasing an extension of the research paradigm in human factors science from "point" (individual human-machine systems) to "2-D plane" (across human-machine systems) and further to the macro environment of STS.

(2). Human-centered AI design philosophy: Emphasis is placed on starting from human needs, values, wisdom, capabilities, and roles. It fully considers the impact of macro-environmental factors such as society, culture, and ethics on the development and use of intelligent systems. This approach addresses new issues in human-machine collaboration and AI ethics within iSTS. User-centered methodologies are employed in system development to ensure effective assistance of intelligent technology in human and organizational decision-making, ensuring human ultimate controllability (Herrmann et al., 2018; Baxter & Sommerville, 2011; Xu, 2019).

(3). Human-machine collaboration based on human-AI teaming: The overlapping portion of the two circles in the social s and intelligent technology subsystem in Figure 4 illustrates the collaborative relationship between these two subsystems. This represents a novel human-machine relationship unique to iSTS (Xu & Ge, 2020). Emphasis is placed on the social and organizational context of iSTS, where intelligent systems are not merely simple tools supporting human tasks and enhancing productivity, as in traditional STS, but also collaborative teammates. In iSTS, interdependent human-AI teams share common societal and organizational goals (Salas et al., 2008). Collaborative interaction, trust, information sharing, and shared decision-making between humans and AI (intelligent agents) are critical factors in the successful development and use of intelligent systems.

(4). Organizational adaptation and redesign: Implementing new AI technologies can alter established work systems, potentially leading some users into difficulties. iSTS emphasizes the redesign of humans and intelligent systems within organizations as a new type of work system, adjusting and optimizing the function and task allocations between humans and machines. This includes redesigning roles, workflows, working environments, and tasks to enhance overall system performance. While introducing intelligent technology, consideration must be given to issues such as job redistribution, fairness, satisfaction, participatory decision-making, and skill development within an organization, thereby enhancing the overall performance of the human-AI system.

(5). Human-AI co-learning: Complex interactions exist between the technical subsystems (hardware, software, intelligent agents, etc.) and human agents (across individual, organizational, and societal levels) within iSTS, transcending the traditional physical boundaries between humans and machines. Intelligent machine agents are a new resource facilitating interaction between the social and technical subsystems in iSTS. This interaction adjusts the behavior of intelligent agents themselves (based on machine learning algorithms, etc.), leading to changes in human usage patterns and expectations (social learning). Simultaneously, the social and technical subsystems of iSTS contain levels of autonomy for humans and intelligent machine agents, reflecting characteristics of co-learning, mutual growth, flexibility, and adaptability. Therefore, iSTS emphasizes that the development and use of intelligent systems need to facilitate human-intelligence co-learning effectively, thereby enhancing the overall capabilities of the human-AI system (Heydari et al., 2019).

(6). Open ecosystem: In the non-intelligent era, the analysis in traditional STS was often a unit with relatively independent boundaries. In the intelligence era, intelligent ecosystems such as the Internet of Things (IoT), smart cities, and intelligent transportation exist within the complex, interdependent iSTS. The development of intelligent technologies will significantly increase the number of intelligent machine agents in iSTS, leading to potential uncertainty and unpredictability in intelligent systems. Features such as human-AI co-learning and potential indeterministic AI output in iSTS introduce dynamic and ambiguous boundaries (van de Poel, 2020). These new features of autonomy present both opportunities for innovative design of intelligent systems and challenges in system design, rules, ethics, cultural values, and more (Hodgson et al., 2013). Therefore, the design, development, and deployment of intelligent systems need to be considered from an open perspective of the human, technical, social, and organizational ecosystem.

#### 3.3.2 Application Analysis

From the perspective of iSTS, human-vehicle co-driving of autonomous vehicles is not merely an engineering and technological project; it is also a sociotechnical project that requires consideration of numerous non-technical factors. For instance, public trust in autonomous vehicles is currently not high (Lee & Kolodge, 2018), and research on human-AI driving ethics primarily focuses on the individual vehicle level (Borenstein et al., 2019). iSTS research on human-vehicle co-driving needs to consider a series of factors, including intelligent autonomous technology, design standards, road traffic infrastructure, regulations and policies, ethical design, traffic rules, intelligent networks/5G, return on investment for enterprises, public trust and acceptance, driver skills, and certification of autonomous driving technology. For example, it involves studying how the manufacturers collaborate with other iSTS members (manufacturers, regulatory bodies, law enforcement, consumers, etc.). There's a need for societal consensus on accident risk mitigation and accountability strategies, such as prioritizing the protection of passengers or pedestrians in emergencies (Chen et al., 2021). It also involves coordinating autonomous driving with existing traffic regulations and implementing the concept of "meaningful human control" in system design. This ensures operators have sufficient situation awareness to make conscious, ethically compliant decisions and control over intelligent systems (Santoni, et al., 2018).

In summary, research on human-vehicle co-driving needs to extend from the paradigm of "human-AI joint cognitive systems" and "human-AI joint cognitive ecosystems" to the paradigm of "intelligent sociotechnical systems". This extension requires interdisciplinary and cross-sector collaboration to develop safe, technologically feasible, publicly trusted, and accepted autonomous vehicles.

#### 3.3.3 Research Prospects

To enrich the iSTS conceptual framework, future research needs to consider two aspects. Methodologically, the iSTS framework emphasizes interdisciplinary teamwork, and the development and use of intelligent systems based on iSTS can utilize human factors science methods such as iterative prototyping and user testing methods to progressively optimize design (Norman & Stappers, 2015). Xu (2022c) has suggested a series of methods based on the contribution of iSTS in various stages of system development, such as ecological research methods (Brown et al., 2017), contextual design (Beyer & Holtzblatt, 1999), anthropological workplace analysis (Hughes et al., 1992), and longitudinal research (Lieberman, 2009). These methods need further refinement and enrichment in future research. Additionally, innovative methods are needed for iSTS research.

In terms of research agenda, iSTS research can be considered from several perspectives. Firstly, share and implement the design philosophy of "human-centered AI" in developing intelligent systems, which will help address more realistic issues for human society. Secondly, there is a need to enrich iSTS theory. For example, how do the social and intelligent technology subsystems co-evolve? How does the interaction between humans and intelligent agents in the social and organizational environment affect human behavior, organizational change, organizational learning, and organizational cognition? How can iSTS design and governance be effectively carried out? Lastly, engage in applications of human factors science based on iSTS. For example, how can the iSTS concept be effectively applied in developing intelligent systems? How can iSTS contribute to the governance of ethical AI (Chopra & Singh, 2018; Fiore, 2020)?

### 3.4 Evolution of the Research Paradigms in Human Factors Science Across Technological Eras

Based on the above analysis, Table 1 summarizes the evolution of the research paradigms in human factors science across different technological eras. Table 1 shows that: (1). Starting from traditional human-machine system theory, the research paradigm of human factors science gradually expanded towards micro and macro directions with the development of technological eras. Technology has driven the expansion of research paradigms, such as the cognitive information processing theory in the computer era and cognitive neuroscience research based on brain activity technology. (2). The research paradigms demonstrate cross-era vitality and enrich human factors research. For example, the traditional paradigm of the static human-machine function allocation was adopted based on the differences and complementarity in capability between humans and machines in the mechanical era; it still holds significance for human factors research in the intelligence era. It helps achieve dynamic human-machine function allocation and human-machine hybrid intelligence in system design by studying the differences and complementarity of human-machine intelligence. (3). The research paradigms exhibit interdisciplinary characteristics, reflecting the developmental needs of human factors science as an interdisciplinary field (e.g., cognitive neuroscience, AI technology). (4). With the expansion of research paradigms, the human factors science community continuously addresses new human factors problems arising from emerging technologies (e.g., human factors issues in computers, intelligent technology). Therefore, the cross-era evolution of the research paradigms in human factors science promotes the development of human factors science, and human factors science research also requires diverse paradigms.

**Table 1: Evolution of the Research Paradigms in Human Factors Science Across Technological Eras**

Microscopic ↕ Macroscopic

| Technological era | Research paradigms | Paradigm description | Representative field or framework | Representative method |
|---|---|---|---|---|
| Computer and intelligence era | Based on human cognitive neural activity | Understand the relationship between the neural mechanism of cognitive processing and work performance in the human-machine environment at the neural level | Neuroergonomics | Brain-computer interface technology and design, EEG measurement, feature analysis and modeling |
| Computer era | Based on human cognitive information processing activities | Understand the relationship between cognitive processing and human performance in the human-machine environment from the perspective of human psychological activities (perception, memory, cognitive load, etc.), and optimize the design of human-machine systems | Engineering psychology | In the human-machine operating environment, human performance measurement (reaction time, error rate, etc.) and subjective evaluation methods are used to evaluate the relationship between human psychological activities and performance and the effectiveness of human-machine system design. |
| Mechanical era | Based on the difference and complementarity in capability between humans and machines | Understand the differentiation and complementarity for optimizing human-machine function and task allocations, and adapting humans to machines | Early work in ergonomics and human factors engineering | Human physical work analysis, time action analysis, human-machine function allocations, etc. |
| Computer era | Human-computer interaction based on machines as auxiliary tools | Achieve machine adaptation to humans, optimized human-computer interaction and user experience based on human-computer interaction technology, design, testing, and implementation | Human-computer interaction | Research and analysis of user psychological models and needs, cognitive modeling of human-computer interaction, interface design conceptualization, and usability testing based on psychological methods |
| Intelligence era | Based on the collaborative relationship between humans and intelligent machine agents as teammates (individual human-computer systems) | Consider intelligent machine agents as teammates collaborating with humans; humans and intelligent machine agents are two cognitive agents in the joint cognitive system; the best overall system performance is achieved through collaboration | Human-AI joint cognitive system based on the human-AI teaming metaphor | Modeling and implementation of human-AI bidirectional and shared situation awareness, psychological models, trust, and decision-making to optimize human-AI interaction and collaboration by leveraging theories such as human-human teamwork and joint cognitive systems |
| Intelligence era | Based on collaboration between cross-human-intelligence systems (multiple joint cognitive systems) | Consider the interaction and collaboration across multi-agent systems (multi-human-AI joint cognitive systems) from the perspective of an intelligent joint cognitive ecosystem. The overall system performance depends on the collaboration and optimized design across joint cognitive systems within a human-AI joint cognitive ecosystem | Human-AI joint cognitive ecosystem | Ecosystem-based modeling, design, and technology, including collaboration across multi-agent systems, group knowledge transfer between multi-agent systems, self-organization and adaptive collaboration, distributed situation awareness, collaborative decision-making, etc. |
| Intelligence era | Based on the optimization between social and intelligent technical subsystems | Achieve the best overall system performance by realizing the optimized interaction and collaboration between intelligent technical subsystems and non-technical subsystems (e.g., humans, organizations, society) | Intelligent sociotechnical systems | Systematic methods, sociotechnical systems methods, work system redesign, organization design, social behavioral sciences, and other interdisciplinary methods. |

Furthermore, we propose three new research paradigms to address the human factors challenges posed by AI technologies in the intelligence era: human-AI joint cognitive systems, human-AI joint cognitive ecosystems, and intelligent sociotechnical systems. Figure 5 illustrates the relationships among these three emerging research paradigms. Based on Table 1 and Figure 5, it is evident that these three emerging paradigms fundamentally adhere to the primary research paradigm of human cognitive information processing in human factors science. Similar to the research paradigms of individuals, groups, and society in behavioral science, these proposed paradigms represent an expansion from the "point" solution (i.e., individual human-AI systems) to the "2-D plane" solution (i.e., across multiple human-AI systems within intelligent ecosystems) and then to the "3-D cube" solution (i.e., the macro-level intelligent sociotechnical systems environment across intelligent technical and non-technical subsystems). This expansion mirrors the new requirements placed on human factors science in the intelligence era, urging solutions beyond isolated traditional research paradigms. Instead, the focus should be on providing comprehensive and systematic solutions for humans and society.

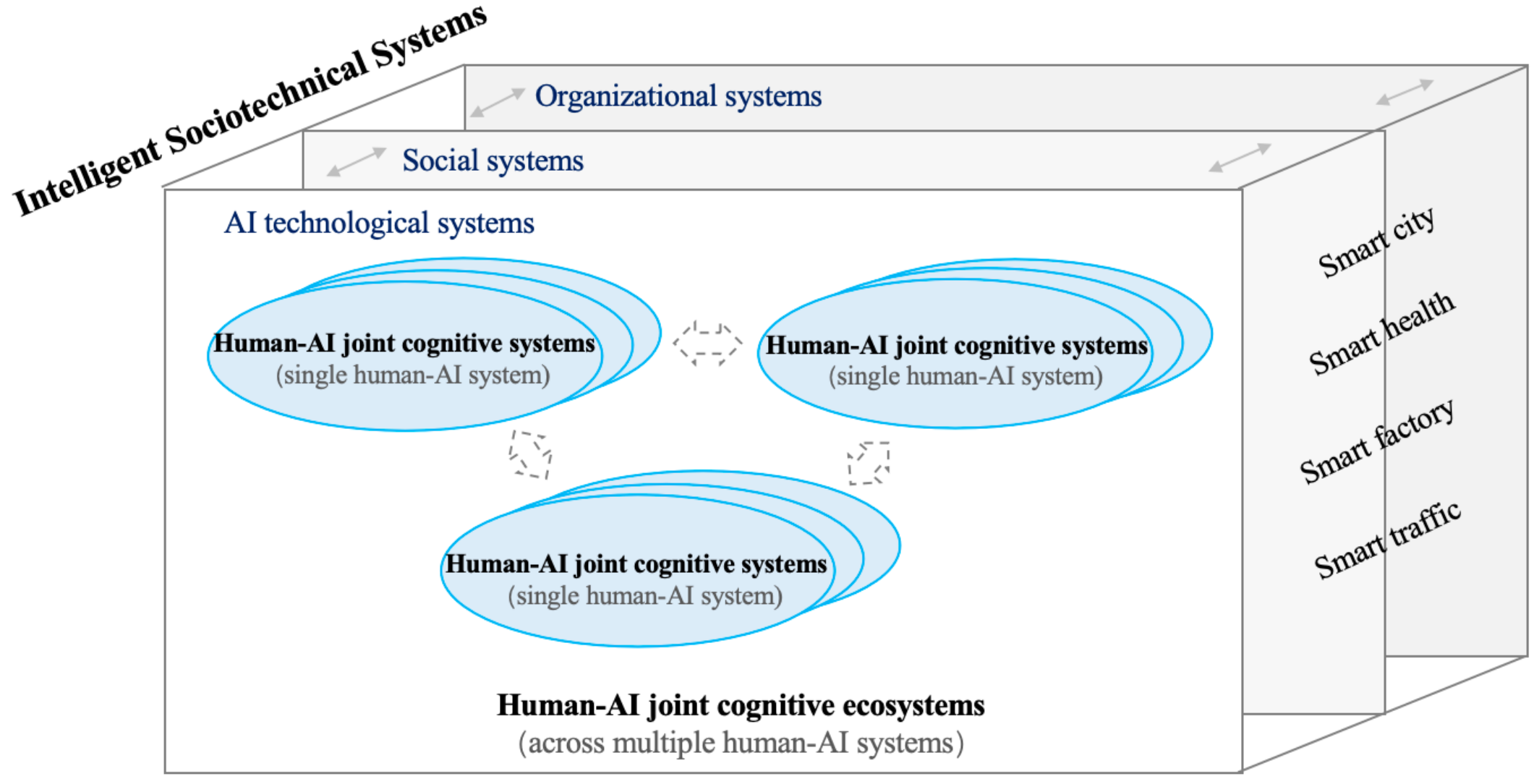

Figure 5 The relationship between three emerging research paradigms of human factors science

## 4 Research Agenda of Human Factors Science in the Intelligence Era

Similar to the cross-era evolution of research paradigms, the transition into the intelligence era brings forth new needs for human factors science research. Drawing upon literature reviews and our recent work (e.g., Xu, 2019, 2021; Xu, 2022, 2022a, 2022b, 2022c; Xu et al., 2021; Xu et al., 2022; Xu, 2023; Xu & Gao, 2023), this paper will highlight the future research agenda of human factors science research based on the transformative characteristics in the intelligence era.

### 4.1 Transition from "Human-Computer Interaction" to "Human-AI Teaming"

"Human-Computer Interaction" (HCI) emerged as an interdisciplinary field during the personal computer (PC) era, focusing on interactions between humans and non-intelligent machine systems. With the transition of machines in the intelligence era to AI-based intelligent systems, a myriad of new features and corresponding human factors issues have arisen, as exemplified in Table 2. Research addressing these human factors issues has been initiated, including studies on "Human-Intelligent System Interaction" (Brill et al., 2018), "Human-Agent Interaction" (Prada & Paiva, 2014), "Human-Autonomy Interaction" (Cummings & Clare, 2015), and "Human-AI Interaction" (Amershi et al., 2019). Although these studies have distinct emphases, they collectively explore interactions between humans and intelligent "machines" (agents). The essence of this interaction is encapsulated in the term "Human-AI Interaction". In this context, Xu, Ge, and Gao (2021) advocate for "Human-AI Interaction" as a novel interdisciplinary research and application domain. Centered around the new transformative characteristics arising from AI technology. Table 2 summarizes the future research agenda of human factors science in the domain of "Human-AI Interaction".

**Table 2: The research agenda of human factors science in the field of "human-AI interaction"**

| Transformative characteristics | New human factors issues from AI technology | Key topics of human factors science research (partially) |
|---|---|---|
| From expected to unexpected machine behavior | • Intelligent systems can bring uncertain machine behavior and unique machine behavior evolution, leading to system output bias (Rahwan et al., 2019)<br>• Existing software testing methods lack consideration of intelligent machine behavior<br>• The behavior of intelligent systems demonstrates characteristics such as evolution and social interaction | • Behavioral science approach to studying machine behavior<br>• Iterative design and user testing methods to avoid system output bias in data collection, training, and algorithm testing (Amershi et al., 2014)<br>• User participatory design, "human-centered" machine learning (Kaluarachchi et al., 2021) |
| From "human-machine interaction" to "human-AI teaming" | • Machines (intelligent agents) also are teammates collaborating with humans<br>• Collaboration between humans and machines<br>• How to model human-machine collaboration (human-machine shared trust, shared situation awareness, mental models, decision-making and control, etc.) | • Theories, methods, etc. based on the human-AI teaming paradigm<br>• Human-AI collaboration theory, model, and team performance evaluation (Bansal et al., 2019) |
| From "human intelligence only" to "human-machine hybrid enhanced intelligence" | • Machines cannot imitate high-order human cognitive abilities, and developing machine intelligence in isolation encounters a bottleneck effect (Zheng et al., 2017)<br>• The integration of human roles into intelligent systems becomes crucial to achieving human-controllable AI (Zanzotto, 2019) | • Cognitive architecture for human-machine hybrid augmented intelligence<br>• Human-multi-agent collaboration systems based on the human-AI joint cognitive ecosystem paradigm<br>• "Human-in-the-loop" and "human-on-the-loop"-based intelligent systems and interaction design<br>• Human high-order cognitive ability models, knowledge representation and graphs |
| From "human-centered automation" to "human-controllable autonomy" | • Humans may lose ultimate control of intelligent autonomous systems<br>• Potential negative impacts of autonomous technology (indeterministic output, etc.) (Kaber, 2018)<br>• Confusion between automation and autonomy technologies can lead to underestimation of the potential negative impacts of autonomous technologies | • Human-computer interaction design paradigm for autonomous technology<br>• The human factors methods of human-controllable autonomy<br>• Human-machine shared autonomous control design<br>• Human-autonomy interaction |
| From "non-intelligent" to "intelligent" human-machine interaction | • How to make intelligent user interfaces more natural<br>• How to effectively design human-AI interaction (Google PAIR, 2019)<br>• Bottleneck effect of human perception ability and cognitive resources in the ubiquitous computing environment (Wang et al., 2014) | • New design paradigms for human-AI interaction and interface design<br>• Multi-channel natural user interface design<br>• Emerging human-machine interaction technology and design (emotional interaction, intention recognition, brain-computer interface, etc.)<br>• Human factors design standards for intelligent technology |
| From "user experience" to "ethical AI" | • New user needs (privacy, ethics, fairness, skill development, decision-making rights, etc.) (IEEE, 2019)<br>• Possible output bias and unexpected results of intelligent systems<br>• Abuse of intelligent systems (discrimination, privacy, etc.)<br>• Lack of traceability and accountability mechanisms for intelligent system failures | • Ethical AI cross-disciplinary design based on human factors science methods<br>• An approach based on the human-AI joint cognitive ecosystem paradigm<br>• An approach based on the intelligent sociotechnical systems paradigm<br>• Meaningful human control (Santoni & van den Hoven , 2018)<br>• Transparency design |
| From "experience-based" to "systematic" interaction design | • Limitations of design methods based on current user experience and usability practices<br>• How to effectively carry out prototype design and usability testing of intelligent systems<br>• Human factors science professionals failed to intervene early in the development of intelligent systems in many cases | • Intelligent system development process based on human factors concepts<br>• AI-based innovative design driven by user experience<br>• Effective intelligent interaction design methods (Holmquist, 2017)<br>• Systematic human factors science methods (Xu et al., 2019) |
| From "physical interaction" to "XR | • New demands for human-machine interaction in extended reality (XR) and metaverse spaces (Shi, 2021) | • Natural human-computer interaction models and technologies in the metaverse space |

| anmetaverse-based interaction" | • The immersion, interactivity, and new experience in the metaverse space<br>• Multimodal continuity and interactive data ambiguity in the metaverse space bring new challenges to interactive intention reasoning. | • Virtualization, remoteness, and multi-mapping relationships of human-computer interaction<br>• Ethics, information presentation, brain-computer fusion, etc., in the metaverse interactive space<br>• Social relationship between human-human and human-AI in the metaverse interactive space |
|---|---|---|

As indicated by Table 2, human-AI interaction research has surpassed the scope of existing human-computer interaction (focusing on human interactions with non-intelligent computing systems). The research and application of human-AI interaction have provided an interdisciplinary platform for human factors science research, facilitating the expansion of research paradigms and the interdisciplinary collaboration to develop "human-centered" intelligent systems collectively.

### 4.2 Transition from "Traditional Human-Computer Interface" to "Intelligent Human-Computer Interface"

Human-computer interface stands as a focal point in human factors science research. In the realm of traditional work based on non-intelligent technologies (e.g., user graphical interfaces), human factors science has aimed to achieve a user-friendly experience through effective user interface design based on approaches such as the "stimulus-response" design scheme and user precise input when interacting with non-intelligent computing systems (Farooq & Grudin, 2016; Xu, 2022a). In the intelligence era, the "traditional human-computer interface" has evolved into the "intelligent human-computer interface", bringing about new features in human-computer interaction across technological eras. For example, the shift from "unidirectional" to "bidirectional" user interfaces, from "usable" to "explainable AI" interfaces, from "simple attributes-based" to "contextualized" interfaces, from "user precise input-based" to "fuzzy reasoning-based" interfaces, and from "interactive" to "collaborative" cognitive interfaces (see Table 3). The major challenge for human factors science is how to address the corresponding human factors issues through these innovative approaches, thereby providing users with a more natural, effective, explainable, and understandable experience. Surrounding these transformative characteristics and emerging human factors issues. Table 3 outlines the future research agenda on human factors science in the "intelligent human-computer interface" domain.

**Table 3: The research agenda of human factors science in the field of "Intelligent Human-Computer Interface"**

| Transformative characteristics | New human factors issues from AI technology | Key topics of human factors science research (partially) |
|---|---|---|
| From "one-way" to "two-way human-machine collaborative" interfaces | • Intelligent systems no longer passively accept user input and produce expected output according to fixed rules.<br>• Intelligent agents can actively sense, capture, and understand users' physiological, cognitive, emotional, intentional, and other states and actively initiate human-computer interaction and offer services. | • Human-computer interaction models based on the human-AI teaming paradigm<br>• Cognitive models of user situation awareness, physiology, cognition, emotion, and intention state |
| From "usable" to "explainable AI" interfaces | • The AI "black box" effect can lead to unexplainable and incomprehensible system outputs (Muelle et al., 2019)<br>• AI "black box" effect raises AI trust issues | • Innovative human-computer interface technology (e.g.visualization) and design<br>• "Human-centered" explainable and understandable AI (Ehsan et al., 2021)<br>• Application of explanatory theories in psychology (Mueller et al., 2019) |
| From "simple attributes" to "contextualized" interfaces | • In addition to simple perceptual attributes of humans, machines, and objects (such as target location and color on the user interface), system inputs also include "contextualized" input targets (such as usage context and user behavior data) | • Modeling of intelligent deduction (e.g., user personal and behavioral patterns) based on interaction context, user behavior, and other data, "<br>• Personalized design suitable for user needs and usage scenarios |
| From "user precise input" to "fuzzy reasoning" interactive interfaces | • User input is not only a single and precise form (such as keyboard, mouse) but also based on multi-modal and fuzzy interaction (e.g., user intention)<br>• Fuzzy interaction-related issues in application scenarios (e.g., random interaction signals and environmental noise) | • Methods and models for inferring user interaction intentions under uncertainty (Yi et al., 2018)<br>• The naturalness and effectiveness of human-computer interaction under fuzzy conditions |
| From "interactive" to "collaborative" | • The user interface must support both human-AI interaction and human-AI teaming | • Effective design paradigms and models of human-AI collaboration-based cognitive interface |
| cognitive interfaces | • Human-machine interfaces that support effective human-machine collaboration | • Interface design standards based on intelligent human-computer interaction<br>• Interaction design that effectively supports human-machine collaboration (e.g., human-machine control handover in emergencies) |

### 4.3 Transition from "Human-Computer Interaction" to "Human-AI Teaming"

Machines based on non-intelligent technologies achieve human-computer interaction through unidirectional, non-sharing mechanisms (i.e., only human trust, situation awareness, and control directed toward the machine) without intelligent complementarity (i.e., relying solely on human intelligence). On the other hand, human-AI teaming, facilitated by intelligent autonomous technologies, enables "bidirectional collaborative" interaction. This form of interaction exhibits characteristics such as bidirectional proactive engagement, information sharing, complementarity, and adaptability between humans and intelligent agents (Xu & Gao, 2020). Therefore, there is a fundamental distinction between traditional human-computer interaction and human-AI teaming (Shively et al., 2018; Brill et al., 2018; Xu, 2021). Research on human-AI teaming is underway (NASEM, 2021; Caldwell et al., 2022; Xu & Dainoff, 2022). Table 4 summarizes the future research agenda of human factors science in the domain of human-AI teaming.

**Table 4: The research agenda of human factors science in the field of "Human-AI Teaming"**

| Key aspects | New human factors issues in human-AI teaming | Key topics of human factors science research (partially) |
|---|---|---|
| Methods and models | • How to quantitatively predict the knowledge structure and interface mechanism of human-AI teaming<br>• How to evaluate the role and performance of intelligent agents in human-AI teams (Demir et al., 2018)<br>• Team performance measurement in complex dynamic scenarios | • Human-AI teaming theories (and methods<br>• Evaluation systems and prediction models for human-AI teaming performance (Kaber, 2018)<br>• Models of an agent's ability to perform expected functions in uncertain scenarios |
| Collaborative process and capabilities | • How human-AI teams collaborate in the long term, function allocation, and goal setting in distributed teams<br>• How do agents coordinate the collaboration of human-AI teams?<br>• Diverse, complex, dynamic, and adaptive collaboration scenarios involved in human-AI teaming (Goodwin et al., 2018) | • Skills for human-AI teaming (e.g., team building, goal setting, communication and coordination, human-AI collaboration language) ( NASEM, 2021)<br>• Effective team processes to support human-AI teaming<br>• The ability of an agent to act as a collaborative coordinator or team resource manager (Wesche & Sonderegger, 2019) |
| Situation awareness | • Human-AI teaming requires teamwork and shared situation awareness<br>• Human-AI teamwork across intelligent systems requires optimized information integration methods<br>• The situation awareness of human-AI teams may be damaged in emergency situations, which is difficult to predict in advance (NASEM, 2021) | • Team-based, distributed, shared situation awareness (Endsley & Jones, 2012)<br>• The relationship between agent self-awareness, awareness of human teammates, and overall team performance (NASEM, 2021)<br>• Situation awareness models to perceive, understand, and predict the collaboration status of human-AI teams |
| Human-machine trust | • Human intelligence trust models and implementation methods need to be rebuilt<br>• Trust research and testing methods need to be restructured | • The impact of human-AI teaming scenarios and goals on trust<br>• Measures of trust in team structure and collaboration<br>• A dynamic model of the evolution of human-AI shared trust |
| Team operations | • How human-AI team members collaborate when sharing system functions<br>• How to realize human-AI teaming management across levels of autonomy<br>• How to implement adaptive operations across levels of autonomy<br>• How to realize dynamic function allocation and collaborative operations across human-AI teams | • Collaborative methods for human-AI teams to share tasks and functions<br>• Methods for human-AI teaming to respond to system autonomous changes under emergency conditions<br>• Requirements for human skill retention and training in human-AI team operations (Roth et al., 2019)<br>• The relationship between human-AI teaming and flexible and autonomous system operations |
| Human-AI co-learning and co-evolving | • The prerequisites required for human-AI teaming (e.g., shared information, knowledge, skills, abilities, goals, and intentions) (van der Bosch et al., 2019)<br>• How human-AI co-learn and co-evolve (e.g., relationships, processes, mechanisms) ( van der Bosch et al., 2019) | • Human-AI team engagement theories and methods (short-term and long-term, task and social participation, participation in dynamic processes) (Madni & Madni, 2018 )<br>• Human-AI team learning models (Schoonderwoerd et al., 2022 )<br>• Team learning, knowledge, and experience sharing model (van der Bosch et al., 2019) |
| | | • Models and methods of human-AI co-evolution ( Döppner et al., 2019) |
| Social factors | • The transfer of social human-human interaction to social human-AI interaction (Schneeberger, 2018)<br>• Lack of understanding of social cognition, social roles, social adaptability, emotions in human-AI teaming | • Social agents in human-AI teams (André et al., 2020)<br>• Group interaction between humans and social agents (André et al., 2020)<br>• Social interaction of human-AI teams (Bendell et al., 2021) |

## 5 Implications of The Emerging Research Paradigms for Human Factors Science

To further evaluate the impact of research paradigms on future human factors science research, we conducted an in-depth analysis of selected key research topics in the intelligence era (see Tables 2 ~ 4). Table 5 presents a summary of these research foci and their association with the research paradigms. The table headers represent diverse research paradigms in human factors science, while the row headers outline pertinent research areas for future exploration. Each cell in the table encapsulates a series of key research questions that need consideration across the research paradigms.

As shown in Table 5, it is evident that, on the one hand, pivotal human factors research topics such as intelligent machine behavior, human-machine collaboration, human-machine hybrid intelligence, ethical AI, intelligent human-machine interaction, and explainable AI require a diverse array of research paradigms for robust support. Established research paradigms in human factors science, such as neuroergonomics, engineering psychology, and human-computer interaction, are still continuing to play pivotal roles. On the other hand, the three emerging research paradigms proposed in this study extend the perspectives, methodologies, and scope of human factors science research in the intelligence era. These paradigms facilitate comprehensive studies in human factors science, offering holistic solutions in the intelligence era.

It is evident that, akin to the evolutionary shifts in human factors research paradigms fostering the discipline's development, the intelligence era necessitates diverse research paradigms. Human factors science research is poised to advance by refining its research paradigms, and in turn, these paradigms will stimulate and enhance research priorities. Therefore, the relationship between research paradigms and research agenda in human factors science has to be mutually influenced and promoted.

## 6 Conclusion

In the context of the emerging concept of human factors science, this paper has delineated the evolving research paradigms in human factors science across technological eras. The intelligence era demands diverse and innovative paradigms for human factors science research. To address these challenges and research needs, we propose three innovative research paradigms for human factors science in the intelligence era based on our conceptual frameworks of human-AI joint cognitive systems, human-AI joint cognitive ecosystems, and intelligent sociotechnical systems. These new research paradigms hold significant implications for the research and application of human factors science.

This paper also looks ahead to crucial research agendas in the areas of "human-AI interaction," "intelligent human-computer interface", and "human-AI teaming". We conclude that diverse and innovative research paradigms in human factors science are conducive to advancing research in the field. Both the research paradigms and research agendas are mutually influential and reinforcing, collectively propelling the development of human factors science.

**Table 5: The Relationship between Research Paradigms and Research Focus of Human Factors Science**

| **Research topics** | **Research paradigms** | | | | | |
|---|---|---|---|---|---|---|
| | Neuroergonomics (neural mechanisms of cognitive processing) | Engineering Psychology (cognitive information processing) | Human-computer interaction (computers as tools) | Human-AI joint cognitive system based on human-AI teaming | Human-AI joint cognitive ecosystems | Intelligent Sociotechnical Systems |
| Intelligent machine behavior | Machine learning algorithm optimization based on human cognitive neural models | Machine learning algorithm optimization based on human information processing models (Leibo et al., 2018) | Optimize machine learning algorithm training and testing to avoid algorithm and behavioral bias based on iterative prototyping and user testing methods | The impact of human-AI collaboration on machine behavior | Machine behavior evolution models, human-machine behavior synergy, and symbiosis theories (Rahwan et al., 2019) | The impact of social environment on machine behavior, machine behavior in social interaction, fairness and ethics of machine behavior, coordination of AI decision-making and organizational decision-making |
| Human-AI teaming | Neural mechanisms in human-AI team collaboration and interaction (Stevens & Galloway, 2019) | Cognitive models such as user perception, emotion, intention, behavior | Human-computer interaction and interface models based on human-AI collaboration | Research on human-AI collaboration models, including human-machine mutual trust, shared situation awareness, mental models, decision-making, etc. | The ecosystem of human-AI teaming, the collaboration between multi-agent systems (Mohanty & Vyas, 2018), the adaptive, self-organizing, and evolutionary mechanism of human-AI systems (Wu, 2020 ) | Human-AI teamwork in a social environment, social interaction between humans and agents, and the impact of social responsibility on human-machine collaboration (Mou & Xu, 2017) |
| Human-machine hybrid augmented intelligence | Research on brain-computer hybrid and brain-computer fusion | Application of advanced human cognitive computing models, knowledge representations, and graphs in realizing human-machine hybrid intelligence | Interaction design based on "human-in-the-loop" hybrid intelligence and human-machine collaborative control (Hu et al., 2020) | Human-AI collaboration in human-machine hybrid augmented intelligence, human-AI complementarity in human-machine hybrid augmented intelligence | Human-machine hybrid intelligence across multiple intelligent systems (Dorri et al., 2018), human-machine-object fusion intelligence (distributed collaborative cognition theories) (Guo & Yu, 2021) | Complementarity and coordination of human-AI teaming in social and organizational environments, distribution of functions and tasks, and setting of human-machine decision-making authority |
| Ethical AI | | Knowledge and methods of ethical AI (Schoenherr, 2022; Chrisle, 2020) | "Meaningful human control" design (autonomous systems) (Santoni & van den Hoven, 2018) | Ethical Issues in human-AI Collaboration | Ecosystem approaches to ethical AI (Stahl, 2021); Compatibility issues (culture, ethics, etc.) across agent systems (ISO, 2020) | Ethical AI issues in intelligent sociotechnical systems, ethical sociotechnical systems ( Chopra & Singh, 2018) |
| Intelligent human-computer interaction | Brain-computer interface technology, design, and application | cognitive models of social and emotional interaction, and intention | New design paradigms and methods of intelligent human-computer interaction, intelligent human- | Cognitive interface design, new design paradigms, and cognitive architecture based on human-AI collaboration | Intelligent human-computer interaction simulation and ecological management, co-evolution of multi-intelligent interactive systems (Döppner | The impact of social, cultural and other factors on intelligent human-computer interaction |
| | | recognition | computer interaction design standards | | et al., 2019) | |
| Explainable AI | Cognitive neuroscience research on explainable AI (Fellous et al., 2019) | Application of psychological explanation theory, cognitive interface models for explainable AI | Innovative human-computer interface technology and design, visualization technology and design | "Human-centered" explainable AI (Ehsan et al., 2019) | Explainable AI problems across intelligent decision-making systems | The relationship between public AI trust and acceptance and AI explainability (Ehsan, 2020), the relationship between explainable AI and culture, user knowledge, decision-making, and ethics |